\documentclass[manuscript,screen]{acmart}
\usepackage{array}
\usepackage{svg}

\newcommand{\bh}[1]{#1}
\newcommand{\kb}[1]{#1}
\AtBeginDocument{%
  }

\setcopyright{acmlicensed}
\copyrightyear{2026}
\acmYear{2026}
\acmDOI{XXXXXXX.XXXXXXX}
\acmConference[Conference acronym 'XX]{Make sure to enter the correct
  conference title from your rights confirmation email}{June 03--05,
  2026}{Woodstock, NY}
\acmISBN{978-1-4503-XXXX-X/2018/06}

\begin{document}

%
\title{Structured State Reconciliation for Human–AI Task Handover}

\author{Kayleigh Bishop}
\email{kayleigh.bishop@colorado.edu}
\affiliation{%
  \institution{University of Colorado Boulder}
  \city{Boulder}
  \state{Colorado}
  \country{USA}
}

\author{Maria P. Stull}
\affiliation{%
  \institution{University of Colorado Boulder}
  \city{Boulder}
  \state{Colorado}
  \country{USA}
}

\author{Breanne Crockett}
\affiliation{%
  \institution{University of Colorado Boulder}
  \city{Boulder}
  \state{Colorado}
  \country{USA}
}

\author{Bradley Hayes}
\affiliation{%
  \institution{University of Colorado Boulder}
  \city{Boulder}
  \state{Colorado}
  \country{USA}
}
\renewcommand{\shortauthors}{Bishop et al.}

\begin{abstract}
  \bh{Task handover requires communicating enough current state for a
  successor to resume work, yet the relevant information is often
  divided between system records and human observations. System records
  can be precise and timestamped but only partially observe the task,
  while human reports capture intent and task knowledge that no log
  contains but are vulnerable to omission and memory error. We present
  a provenance-aware pipeline that converts task telemetry and
  human-authored reports into a shared typed task-state representation,
  aligns and reconciles their facts, detects conflicts, and generates
  structured handover reports. We evaluate the approach on 13 paired
  task states collected in a controlled spatial multitask environment,
  using task-grounded metrics that estimate the state-reconstruction
  cost a report would spare a hypothetical recipient and the
  misinformation burden it would impose. Reconciling both sources
  preserved greater estimated task-state utility than either the user
  report or telemetry alone. Relative to a direct end-to-end LLM given
  the same inputs, structured reconciliation maintained comparable
  estimated utility while incurring substantially less misinformation,
  and task-aware rendering retained utility more efficiently (per
  token) than exhaustive rendering. An exploratory content analysis
  further shows that human reports contain substantial strategic
  knowledge that lies outside state-focused metrics. These results
  support provenance-aware state reconciliation as a design pattern for
  safer AI-assisted handover; recipient-side task performance is not
  directly measured and remains an important validation step.}
\end{abstract}




\maketitle

\section{Introduction}

Task handover, the transfer of responsibility for an ongoing task from one agent to another, is among the most frequent sites of critical error in high-stakes domains. In medical settings, communication failures at care transition are a leading contributor to preventable adverse medical events; incomplete or inaccurate transfer of patient state information leads to missed treatments, duplicated interventions, and delayed responses to deterioration \cite{starmer_changes_2014, lyhne_towards_2012}. Analogous cascading failures resulting from handover miscommunication have been documented in aviation, chemical processing, and emergency response \cite{newman_it_2023, noauthor_investigation_2019, manias_communication_2015}.  Standardization tools and mnemonics like I-PASS or SBAR (both utilized in the healthcare domain) aim to reduce the cognitive burden of handover by imposing a consistent structure to handover reports, reducing reliance on ad hoc recall. While these interventions have shown measurable reductions in reported errors \cite{starmer_changes_2014, muller_impact_2018}, the onus remains on the outgoing agent to identify, retain, and articulate task-relevant information, all while managing high workloads, frequent interruptions, and the natural decay of contextual detail over time. Researchers have highlighted the ongoing challenges of addressing the dynamic needs and complex decision support demanded by fast-paced or unpredictable environments, and the outstanding need for effective information support systems in handover \cite{ambe_patient_2025}. \bh{Recent work has begun to apply natural language processing and large language models (LLMs) to handover documentation itself: benchmarking speech recognition and information extraction over nursing handover \cite{suominen_benchmarking_2015}, generating and evaluating emergency-medicine handoff notes \cite{hartman_developing_2024, genes_generative_2025}, and deploying LLM-assisted nursing handover documentation across hospitals \cite{chen_integrating_2026}. These studies establish that LLMs can draft handover documentation and reduce documentation burden, but they generally synthesize a single record stream and evaluate the quality of the resulting draft. A different and comparatively underexplored problem motivates the present work: at handover time, the information a successor needs is typically divided between two \emph{partial and potentially conflicting} accounts (system records and the outgoing worker's own report), and an effective support system must reconcile them, preserve where each piece of information came from, and control the misinformation it passes on.}

The difficulty of handover lies in the inherent challenge of situation awareness (SA) transfer across an agent boundary. During a task, a worker's SA is continuously updated as they perceive relevant state information, integrate it into a coherent model of the situation, and project it forward \cite{endsley_toward_1995}. This awareness is inherently dynamic and perishable; interruptions and multitasking cause memory traces for lower-salience goals to decay, requiring active recovery through re-engagement with environmental cues \cite{gartenberg_situation_2014}. At handover time, however, the incoming agent must acquire SA from the outgoing agent's report, or risk losing valuable time to manually re-acquiring key state information. Thus the handover report (typically a written artifact, though often accompanied by supplementary verbal exchange \cite{bosua_fostering_2015, patterson_handoff_2004}) must often serve as the sole external scaffold for SA reconstruction. \bh{This framing highlights one necessary component of handover quality: whatever else a report conveys, the incoming agent must be able to reconstruct a sufficiently accurate model of the current task state from an incomplete externalization of the outgoing agent's awareness. We focus on this recoverable-state component throughout, while recognizing that interactive communication, strategic guidance, and common-ground negotiation also contribute to successful handover \cite{resnick_grounding_1991, patterson_handoff_2004}.}

\bh{In instrumented workplaces, the raw material for reconstructing task state is divided between two sources with complementary strengths and failure modes. System records (telemetry, event logs, electronic documentation) are precise and timestamped but capture only what has been instrumented, omitting dialogue, intent, and contextual observations. Human accounts capture exactly this uninstrumented content, including strategy and anticipated next steps \cite{toccafondi_collaborative_2012, ambe_patient_2025}, but are vulnerable to omission and memory error \cite{ye_handover_2007, desmedt_clinical_2021}. Effective handover assistance must therefore \emph{reconcile} two partial accounts of the same underlying state, not merely summarize whichever one is available.}

\bh{An obvious approach is to place both sources in an LLM's context window and request a summary. Direct synthesis of this kind, however, leaves several hard problems implicit: the sources differ in reliability; they can contradict each other; event records mix stale and current facts; and generation can introduce content supported by neither source \cite{ji_survey_2023}. Providing source material does not by itself guarantee grounded output \cite{stolfo_groundedness_2024}, and clinician-reviewed evaluations of LLM-drafted handoff summaries continue to report omissions and mischaracterizations that require human review \cite{hartman_developing_2024, genes_generative_2025}. A free-form summary also carries no explicit provenance: the recipient cannot distinguish statements corroborated by system records from statements resting solely on the outgoing worker's memory, or from statements the two sources actively dispute \cite{gao_enabling_2023, ernst_where_2025}.}

\bh{We present a provenance-aware pipeline that treats handover report generation as \emph{structured state reconciliation}. The pipeline converts system telemetry and the outgoing worker's free-text report into a shared typed knowledge representation, aligns entities and facts across the two sources, detects conflicts, reconciles temporal state so that the representation reflects current rather than historical conditions, and preserves the provenance of every fact. A constrained LLM rendering step then realizes the reconciled state as a report. Because not all facts are equally costly to rediscover, a task-aware rendering variant foregrounds outstanding needs and the locations required to satisfy them, while an exhaustive variant reports all available state \cite{yang_effect_2013, rosario_using_2022}.}

\bh{We evaluate the approach on 13 paired task states collected in a controlled spatial multitask handover environment inspired by information-work challenges in healthcare \cite{potter_analysis_2005, patterson_handoff_2004, randell_importance_2011}, in which participants made partial task progress under time pressure and interruption before writing a handover report. Our evaluation is deliberately \emph{artifact-level}: we score each report's content against the ground-truth task state under an explicit search-cost model, estimating (i) the state-reconstruction effort the report would spare a hypothetical recipient and (ii) the misinformation burden it would impose. We do not observe a recipient resuming the task; downstream recipient performance is the motivating application of this work, not its measured outcome.}

\bh{Our study is organized around three research questions and one exploratory question:
\begin{itemize}
    \item \textbf{RQ1 (Source complementarity).} Does reconciling human-authored reports with system telemetry improve estimated recoverable task-state utility relative to either source alone?
    \item \textbf{RQ2 (Value of explicit reconciliation).} Does explicit structured state reconciliation reduce the misinformation burden of generated reports relative to direct end-to-end LLM synthesis over the same source inputs?
    \item \textbf{RQ3 (Task-aware rendering).} Does task-aware rendering improve task-value density (estimated utility per token) relative to exhaustive rendering while preserving comparable estimated task-state utility?
    \item \textbf{Exploratory.} What information do human-authored reports contribute that is not represented by state-recovery metrics?
\end{itemize}}

\bh{This work makes three contributions:
\begin{itemize}
    \item A provenance-aware state reconciliation pipeline for task handover that converts system telemetry and a human-authored report into a shared typed representation, aligns entities and facts, detects and preserves conflicts, reconciles temporal state, and constrains LLM generation to the reconciled result. Unlike prior handover documentation systems that synthesize a single record stream, the pipeline treats the two accounts as jointly partial and mutually correcting.
    \item A task-grounded, artifact-level evaluation framework that scores handover reports against ground-truth task state in cost-denominated units, separating estimated task-state utility from misinformation burden rather than treating all extracted facts as equally valuable. The framework is task-specific by design; its purpose is to make report quality objectively measurable for controlled comparison.
    \item Empirical evidence from 13 paired task states that (i) reconciling both sources preserves more estimated task-state utility than either source alone, (ii) explicit reconciliation substantially reduces misinformation relative to an end-to-end LLM given identical inputs, at comparable estimated utility, and (iii) task-aware rendering achieves a better utility--length tradeoff than exhaustive rendering. An exploratory content analysis additionally documents substantial strategic knowledge in human reports that state-focused metrics do not credit, informing how such systems should divide labor between human and machine.
\end{itemize}}
\section{Background and Related Work}
To design and evaluate an intervention for handover support, we drew on existing research across human factors, computing for collaborative work, and explainable AI. 

\subsection{Handover as collaborative knowledge transfer and information work}
Prior research characterizes handover as a collaborative and context-dependent process in which responsibility must be transferred under severe temporal and cognitive constraints. Bosua and Venkitachalam \cite{bosua_fostering_2015} collected case studies of handover practices from multiple industries, including manufacturing and customer service, and examine them through the lens of their use of "boundary objects": artifacts, such as notes or written reports, that bridge the communicative boundary between collaborators \cite{pawlowski_supporting_2000}. In their work, the authors identified the most common boundary object as the handover sheet. This is typically a standard form that workers use to document relevant information that collaborators should be aware of to ensure continuity of progress. Handover sheets provide a much-needed infrastructure to scaffold knowledge transfer. Workplaces that lacked such an artifact reported lost productivity due to both repeated work and a lack of shared procedural knowledge; one CEO complained that "...[employees] keep solving the problems over and over again during shifts" \cite{pothier_pilot_2005, bosua_fostering_2015}.

Further work, particularly in the healthcare domain, has investigated the role of information work in ensuring task continuity. Information work refers to the process of seeking, receiving, and sharing information, and thus encompasses both handover itself and the myriad steps of information gathering, synthesis, and interpretation that it requires \cite{wilson_human_2000}. Recent work has identified the fragmented nature of information support tools in settings such as emergency departments, and highlighted the necessity of integrating, retaining, and processing patient information dynamically \cite{yang_clinical_2011, ambe_patient_2025}. Yang et al. identified information access cost (IAC) as a key determinant of junior doctors' pre-handover information work; clinicians were more likely to rely on memory alone for patient information that took more time to verify, resulting in more errors during handover preparation \cite{yang_effect_2013}.  This manual burden, in which synthesis and integration must be performed anew by a time-constrained sender or recipient at significant risk of information loss, is where automated assistance offers leverage. It further motivates our decision to operationalize the value of handover content via the information work it saves the recipient, while exploring how automated synthesis of system- and human-authored information can reduce that burden.

\subsection{Situation awareness and distributed cognition at handover}

Endsley's three-level model characterizes situation awareness (SA) as the perception of elements in the environment, the comprehension of their meaning in relation to current goals, and the projection of their future states \cite{endsley_toward_1995}. Applied to handover, this framing clarifies that a complete report is not simply an inventory of observed facts (perception) but must also convey their task-relevant significance (comprehension) and their likely trajectory (projection); for instance, not only that a door is locked, but it blocks progress towards a current goal, and the recipient should expect to need a specific key soon. This distinction between raw state and its interpreted significance has two consequences for our own framework. First, it points to why conveying significance efficiently, and not merely completely, matters at \bh{handover}: a report that lists every perceived fact without signaling which are comprehension- or projection-relevant imposes its own cost on the recipient, who must then perform that interpretive work themselves. \bh{This motivates our task-value density measure (see Evaluation), which evaluates reports not just on the estimated task-state utility they preserve, but also on how economically that utility is conveyed per unit of report length.} Second, it makes clear the need to parse strictly state-related information (perception) from procedural knowledge (projection) information, which we address via our taxonomy of content in handover reports.

In practice, however, SA is rarely held exclusively by a single person; in team settings it must be distributed and reconciled across individuals with different access to information. Clark et al. reviewed handover tools and techniques through this distributed SA lens across healthcare, aviation, energy, and other high-risk industries, cataloguing nineteen distinct techniques in use \cite{clark_identified_2019}. Their comparison of domain-specific mnemonics is instructive: healthcare's SBAR is built around a holistic, interpretive account of patient status, while air traffic control mnemonics such as PRAWNS convey discrete, descriptive facts (e.g., runway in use, barometric pressure) and rely on the operator's own schemata to interpret their significance \cite{clark_identified_2019}. This split between content meant to convey interpreted significance and content meant to convey state data recurs informally across the domains they survey.

Individual studies within this space have gone further, formally modeling which information matters most to distributed SA at \bh{handover}. Rosario et al. used Bayesian belief networks to identify which process variables most affect a recipient's distributed SA in shift changeovers, and should therefore be prioritized in verbal handover updates \cite{rosario_using_2022}. Their approach demonstrated that not all information is equally consequential to reconstruct SA after a \bh{handover}, and that principled prioritization is possible. This finding directly motivates our information access cost metric, which likewise treats facts as differentially costly to omit rather than uniformly weighted. Their method's reliance on a finite, hand-specified set of process variables and hand-analyzed example reports, however, limits its applicability to less structured or more open-ended task domains; our pipeline's use of an LLM-derived fact ontology is intended to relax this constraint.

A complementary line of work addresses the process by which two parties come to a shared situational understanding. Clark and Brennan's theory of common ground describes communication as a joint process of establishing and updating mutually known information, with participants continuously inferring what can safely be assumed shared versus what must be made explicit \cite{resnick_grounding_1991}. This offers a useful lens on why knowledge-transfer content is so pervasive in the user-generated reports we collect. Senders are not simply reporting state, but making implicit judgments about what a recipient with full state visibility would still lack (i.e., what cannot be inferred from telemetry alone). Where SA theory describes what a mental model must contain, common ground theory explains why some of that content resists formalization into discrete, verifiable facts describing task state, and is instead conveyed as heuristic or experiential guidance.

\subsection{Standardization interventions and evaluation}
Efforts to improve handover in high-stakes domains have largely centered on standardizing what gets communicated, rather than on measuring communication quality directly. In healthcare, the now widely-adopted I-PASS handoff program introduced a structured mnemonic and standardized verbal and written format for shift-change reporting across multiple pediatric residency programs \cite{johnson_impact_2016}, and similar benefits have been reported for other standardization protocols such as SBAR \cite{starmer_changes_2014}. Implementation of these programs was associated with significant reductions in medical errors and preventable adverse events, demonstrating that structuring what a sender is prompted to include can meaningfully improve handover outcomes in practice. \bh{Notably, evaluation of these protocols has been largely indirect: standardized, reliable measurement tools for handoff quality itself have remained elusive \cite{patterson_handoffs_2010}, so effectiveness has typically been measured downstream, via the rate of medical errors following implementation, rather than via any property of the reports or the receiving clinician's resulting task state. This leaves open a methodological question that standardization efforts do not by themselves answer: given two handover reports of the same task state, which preserves more of the information a recipient would otherwise have to reconstruct at a cost?}

This absence of a direct evaluation method has not gone entirely unaddressed. O'Connel et al. developed and psychometrically validated the Handover Evaluation Scale, a 14-item self-report instrument spanning quality of information, interaction and support, and efficiency subscales, intended to monitor handover practices across healthcare organizations \cite{oconnell_construct_2014}. This instrument is valuable for organizational quality assurance, but it measures participants' subjective perception of handover effectiveness at the level of an overall process, rather than providing an objective, artifact-level measure of how much task-relevant information a specific report preserves or omits; nor is it designed for comparison between individual reports or reporting strategies, which our task-aware and baseline conditions require. \bh{The methodological need therefore persists at the level our study operates on: an artifact-level, task-grounded way to compare candidate reports of the same underlying state. Consistent with Patterson and Wears's observation that general-purpose handoff measurement remains elusive \cite{patterson_handoffs_2010}, we position our evaluation framework as a task-specific instrument for controlled comparison, not a resolution of the general measurement problem.}

Similar dynamics appear outside healthcare. \bh{Industrial and manufacturing settings have adopted a range of AI-assisted handover tools (documented largely in industry materials rather than peer-reviewed evaluations, so we describe them here as practice context rather than as validated evidence), which fall broadly into two categories.} The first automates system-side information work to improve situation awareness: digital shift logs auto-populate from PLC/SCADA data, dashboards flag deviations and route exceptions to responsible teams, and scheduling tools optimize shift staffing to smooth transitions \cite{virtualworkforceAssistantManufacturing, sysgenproManufacturingAutomation}. The second targets the human side, echoing the aims of standardization protocols in healthcare: voice-to-text logbooks, multimedia entries, and generative drafting tools aim to capture a worker's account of a shift more consistently and richly \cite{jrsinnovationShiftHandover, leewayhertzCasesManufacturing}. Both categories aim to improve on unaided manual handover, but neither treats the system-derived account and the human-authored account as two partial, potentially conflicting models of task state to be reconciled with one another; each instead automates or enriches one side of the handover in isolation, and, as in the clinical case, reports benefits in terms of speed and completeness of capture rather than any measure of what the recipient still needs to reconstruct after reading the result.

\bh{Across clinical standardization, self-report instruments, and industrial tooling alike, a common pattern recurs: interventions standardize or enrich one account of the shift in isolation, and report their benefits in terms of capture speed and completeness. The relationship between the system-derived account and the human-authored account, two partial and potentially conflicting descriptions of the same task state, is generally left implicit. Our evaluation framework instead scores handover artifacts directly against ground-truth task state in a controlled environment, and our pipeline treats telemetry and user report as complementary inputs to a single reconciled, provenance-preserving representation.}

\subsection{\texorpdfstring{\bh{AI and NLP for handover documentation}}{AI and NLP for handover documentation}}

\bh{Applying language technology to handover documentation is not itself new. Suominen et al. established benchmark data and methods for clinical speech recognition and information extraction over nursing shift-change handover, framing the conversion of verbal handover into structured, filterable documentation as an NLP task \cite{suominen_benchmarking_2015}. More recently, LLMs have been applied to drafting handoff documents directly. Hartman et al. developed and evaluated LLM-generated emergency-medicine handoff notes against physician-written notes, coupling automated completeness metrics with a clinical evaluation of usefulness and patient-safety implications, and found that automated notes could be useful but required careful safety evaluation before deployment \cite{hartman_developing_2024}. Genes et al. had emergency-department providers review generative-AI handoff summaries, documenting omissions and mischaracterizations that underscore the continued need for human review \cite{genes_generative_2025}. Chen et al. reported a multi-hospital deployment of LLM-assisted nursing handover documentation, achieving documentation-time savings under constrained prompts and mandatory nurse verification of generated content \cite{chen_integrating_2026}. In HCI, Solano-Kamaiko et al. investigated how conversational AI might facilitate coordination and shift-style information sharing among home care workers and family caregivers, highlighting design tensions around trust and information asymmetry in care networks \cite{solano-kamaiko_sharing_2026}.}

\bh{Collectively, these studies establish the feasibility and documentation benefits of AI-drafted handover artifacts. However, they generally synthesize a single record stream (e.g., the electronic record) or evaluate the quality of a drafted note against clinician judgment. The present work examines a complementary problem that these systems leave implicit: explicitly reconciling a system-derived account and a human-authored account that differ in observability and reliability, preserving the provenance of each fact through generation, and evaluating the resulting artifact against ground-truth task state rather than against reviewer impressions alone.}

\subsection{\texorpdfstring{\bh{Grounded generation, provenance, and structured intermediate representations}}{Grounded generation, provenance, and structured intermediate representations}}

\bh{Our pipeline design is also informed by work on factual reliability in language generation. Surveys of hallucination in natural language generation distinguish output that is unfaithful to its source material from output that is factually incorrect about the world \cite{ji_survey_2023}, a distinction our misinformation analysis inherits. Merely providing source material does not guarantee grounded output: even retrieval-augmented long-form generation frequently produces statements unsupported by the retrieved context \cite{stolfo_groundedness_2024, lewis_retrieval_2020}. One mitigation strategy is attribution: enabling models to generate text with citations to supporting evidence \cite{gao_enabling_2023}, and attributing even erroneous sentences to their most likely source so that errors can be located and repaired \cite{ernst_where_2025}. In clinical summarization specifically, logic-controlled generation systems constrain discharge-summary content by explicit rules and retain source attribution \cite{yuan_lcds_2025}. Our approach applies these ideas to handover: a typed, provenance-preserving intermediate representation with explicit conflict detection serves as a reliability mechanism that reduces the generation step to constrained realization of already-reconciled state, consistent with broader evidence that structured intermediate representations improve LLM performance on complex tasks \cite{delorenzo_abstractions--thought_2025}. We treat the graph structure itself as incidental; what matters are the properties it enforces: typed facts, provenance, temporal status, and inspectable conflicts.}

\subsection{\texorpdfstring{\bh{Relationship to model reconciliation}}{Relationship to model reconciliation}}

Prior work in explainable AI has formulated explanation generation as a problem of reconciling the mental task models of agents and humans \cite{chakraborti_plan_2017}. Under this formulation, an explanation system maintains a second-order model of the recipient's current understanding of task state, and produces explanations as targeted "model updates" that optimize the recipient's expected task performance against the cost of communicating the update \cite{sreedharan_foundations_2021, tabrez_one-shot_2021}. A limiting assumption in this line of work is that both the recipient's policy and the true, static state of the world are known in advance. This assumption does not hold in handover: the dynamic nature of real-world workplaces means it is often impossible to determine a priori which information will prove relevant to the recipient \cite{ambe_patient_2025}. Clinical practice reflects a similar tension: the SBAR and I-PASS handover mnemonics, among the most widely used in healthcare, explicitly separate factual "Situation/Background" reporting from forward-looking "Recommendations" and "Contingency planning" \cite{starmer_changes_2014, muller_impact_2018}, echoing the state/knowledge distinction we formalize in our own taxonomy. \bh{For this reason, we deliberately formulate handover support as provenance-aware \emph{state reconciliation} rather than model reconciliation, and we regard this as the correct formulation for the setting rather than a simplification of it. At handover time the recipient's belief state is unavailable in principle: the incoming worker may not yet be identified, brings unknown prior knowledge, and cannot be queried before the report must exist. A system built around a second-order model of that recipient would rest on an unverifiable estimate, and its outputs could only be judged against further assumptions. Reconciling the two evidence sources that demonstrably exist (telemetry and the outgoing worker's report) instead anchors every retained fact to concrete, checkable evidence, which is what makes our approach objectively measurable: each reconciled fact and each rendered claim can be scored against ground-truth task state, and our evaluation does exactly this. We retain classical model reconciliation as motivation for treating reports as targeted updates rather than transcripts, but the implemented and evaluated object is the reconciled state estimate.}

\bh{The comparisons in our evaluation map directly onto the research questions introduced in Section 1. RQ1 (source complementarity) contrasts the fused pipeline against user-report-only and telemetry-only inputs. RQ2 (value of explicit reconciliation) contrasts the structured pipeline against a direct end-to-end LLM given identical source inputs. RQ3 (task-aware rendering) contrasts task-aware against exhaustive realization of the same reconciled state. The accompanying content analysis of what human reports contribute beyond task state is exploratory: we state no directional hypothesis about whether participants emphasize strategic knowledge over state.}
\section{\texorpdfstring{\bh{Provenance-Aware Task-State Reconciliation}}{Provenance-Aware Task-State Reconciliation}}

\begin{figure*}
    \centering
    \includegraphics[width=1.0\linewidth]{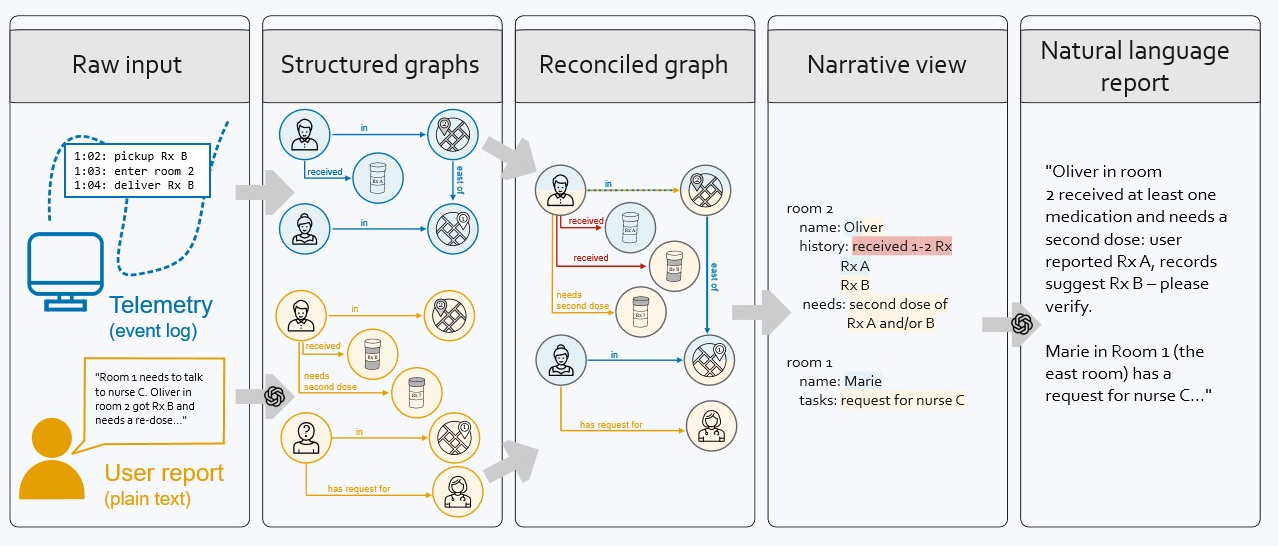}
    \caption{\bh{Provenance-aware state reconciliation pipeline. From left to right: (1) Raw input: source evidence (timestamped telemetry events and sentences from the human-authored report); (2) Structured graphs: source-specific typed facts extracted from each source; (3) Reconciled graph: alignment and conflict detection, classifying each fact as corroborated, telemetry-only, human-only, or conflicting; (4) Narrative view: temporal reconciliation, which prunes superseded facts to yield a current-state estimate with provenance retained,} \kb{and conversion to top-down entity-centric view;} and (5) Natural language report: constrained report realization. The pipeline's contribution is not conversion of text into a graph per se, but preservation of where each fact came from, explicit detection of disagreement, computation of \emph{current} (rather than historical) state, and constrained generation over the result.}
    \Description{A five-column pipeline diagram showing telemetry events and human report text being converted to typed facts, aligned and checked for conflicts, reconciled into a current task state with provenance labels, and rendered as a structured handover report.}
    \label{fig:pipeline}
\end{figure*}

\bh{We characterize handover support as a \emph{provenance-aware state reconciliation} problem. Let $O_s$ denote the timestamped observations logged by the task environment (the system trace), and let $R_h$ denote the free-text report written by the outgoing worker (the ``sender''). Each source induces a set of candidate task-state facts; for each fact $f$ we track its provenance $P(f)$ (which source(s) support it), its temporal position where available, and its support status after cross-source comparison (corroborated, single-source, or conflicting). The system's objective is deliberately modest: construct a current task-state estimate that (i) retains supported, task-relevant facts from both sources, (ii) distinguishes source and uncertainty for every retained fact, and (iii) minimizes unsupported content in the realized report $Y$. Facts differ in how costly they are for a recipient to re-acquire (e.g., re-discovering the location of a person or item requires physical search), and this asymmetry motivates both our task-aware rendering condition and our evaluation metric (Section 5). We emphasize the boundary of this formulation, and that it is drawn deliberately: we do not model the recipient's beliefs or policy, because at handover time these are unobservable and often the recipient is not yet known, and any formulation anchored to them would make the system's correctness unverifiable. Anchoring the formulation to world state instead keeps every output claim objectively checkable, and the realized report is evaluated accordingly, against ground-truth task state under an explicit search-cost model.}

Our system generates structured handover reports by combining two complementary information sources (a system trace logged by the task environment and a free-text verbal report written by the outgoing worker) into a unified knowledge representation. The system trace is objective and verifiable, but intentionally limited to capture discrete, instrumentable events (such as state transitions and agent-environment interactions) while excluding information that resists structured logging, such as dialogue content, agent intent, or contextual reasoning. An LLM then renders this representation into a natural-language report for the incoming worker. We describe each stage of this pipeline below and then present the ablation conditions used to isolate the contribution of each component.

\subsection{Knowledge graph schema}
We represent task state as a typed knowledge graph whose nodes are entities (people, items, locations) and whose edges are structured facts. Our schema, defined using Pydantic models, supports three fact types:

\begin{enumerate}
    \item Location facts: anchor an entity to a named room (e.g., "Lily is in Room 1").
    \item Relation facts: capture task-relevant predicates using a controlled set of relation types (e.g. \textit{needs-medication, received-medication, awaiting-message})
    \item Spatial facts: encode directional references that cannot yet be resolved to a specific room or entity (e.g., "someone in the west hall is waiting on a message").
\end{enumerate}

Entity arguments can be named (a specific person or item) or existential (an underspecified reference such as "someone" or "a message"), with an optional spatial constraint. This distinction allows the schema to faithfully represent the varying specificity typical of verbal reports.

\subsection{Semantic graph construction}
Both inputs are converted to semantic graphs in a shared format for the purposes of later comparison. For our task, system telemetry uses a consistent text template for world events that can be deterministically converted into the graph format.

The task environment logs timestamped events (room transitions, character interactions, medication pickups) to a text file (the system trace). A rule-based parser extracts facts from these events using pattern matching, maintaining a running state tracker for player location and most recently interacted patients to resolve implicit references. This transform is fully deterministic and produces a complete record of every \textit{observable} (to the system) action the outgoing player performed.

User written notes are first converted via LLM to a templated domain-specific language (DSL), extracting state-related facts to simple text strings (e.g. "Oliver is in room 2"). The DSL is constrained by a set of fixed templates and few-shot examples that enforce consistent formatting. Second, a deterministic parser converts each DSL line into a typed fact, handling entity normalization, directional parsing, and existential argument detection.

\bh{This two-stage extraction design (LLM for surface-form normalization followed by deterministic parsing) confines the language model's role to information extraction. We note the precise reproducibility boundary: all transformations \emph{after} each LLM-produced structured output are deterministic, but the pipeline is not LLM-free end to end, since both this normalization step and the existential-reference resolution during entity alignment (below) invoke a model. Model identifiers, prompts, and decoding settings for every LLM call are reported in Section 5 and the supplementary material.}

\subsection{Graph alignment and merging}
The resulting graphs from the previous step (referred to as the "system graph" and "user graph" respectively) are able to be directly compared to identify novel facts, omissions, and potential contradictions via two alignment steps. 

\textit{Entity alignment} maps named entities across the two graphs using string normalization and an alias table, then resolves existential arguments (e.g., "someone to the north") against the telemetry entity set using an LLM call, classifying each as resolved, ambiguous, or unresolvable. Entities in a given graph described with partial and/or existential descriptors (e.g. "the patient in room 1" or "someone in the west wing") are resolved to their matching exact referent in the corresponding graph, if such a match can be found and there is sufficient information to resolve it.

\textit{Fact alignment} matches facts on identity (same entity and predicate) and compares them field-by-field. The result classifies each report fact as confirmed (matching between the two graphs), novel (present only in the report), or conflicting (matching entity/predicate but differing on a value). Contradictions are flagged only in cases where two facts are incompatible (e.g. a person cannot be in multiple rooms at once).

Facts from both graphs are added to a final graph with a flag to indicate their origin and classification, including a record of detected conflicts.

\subsection{\texorpdfstring{\bh{Provenance and conflict representation}}{Provenance and conflict representation}}
\bh{Every fact admitted to the merged graph carries an explicit provenance record: whether it is supported by telemetry, by the user report, or corroborated by both, together with its alignment classification (confirmed, novel, or conflicting) and a record of any detected conflict. Disagreement is not silently collapsed at merge time: when the two sources assert incompatible values, both facts are retained and flagged, so that downstream stages can reason about the conflict as a first-class artifact rather than receiving a single silently-chosen value. Facts supported only by the user report are, by the observability design of our environment, unverifiable from telemetry; they are retained but remain distinguishable from corroborated facts throughout the pipeline. In the current implementation, provenance and conflict records are available to the report generator through the narrative view (below), but are not rendered as explicit source labels in the final report prose; we return to the implications of this design choice in the Discussion and Limitations.}

\subsection{State reconciliation}
A reconciliation pass replays the logical consequences of event-type facts over the declarative state. For example, \textit{(lily, needs, medication 1)} is pruned if a later-timestamped \textit{(lily, received, medication 1)} was added. This ensures the graph reflects the current outstanding task state rather than a cumulative event log, a critical distinction for generating actionable handover reports. \bh{The predicate vocabulary of these rules is domain-specific (e.g., \textit{needs} $\rightarrow$ \textit{received}, \textit{awaiting-message} $\rightarrow$ \textit{delivered}, location superseded by later movement), but the underlying pattern of later-timestamped event facts superseding earlier declarative state is general, and porting the pipeline to a new domain requires specifying the analogous supersession rules for that domain's predicates.}

\subsection{Narrative view construction for top-down view}
The reconciled final graph from the previous step is restructured to a top-down view, where state facts are listed by location and relevant entity. It also collects any unanchored or miscellaneous facts that could not be resolved to a particular entity. Facts are converted back from graph triples to natural language strings. The resulting "narrative view" preserves the context, provenance and confidence data from previous steps while converting key semantic data into a format more suitable for LLM input. 

While earlier versions of the pipeline omitted this step and used the reconciled graph directly in the LLM prompt, we found that restructuring the graph around rooms and task requirements helped performance by providing the LLM with information that is already organized for spatial reasoning and task prioritization, rather than requiring the model to navigate a flat list of heterogeneous facts.

\subsection{Report generation}
While our system's task modeling and reconciliation provide an organized view of task history and information, these steps do not address the problem of selecting key information to include in \bh{the report}, or conveying that information in a human-readable way. To this end we leverage an LLM to make inferences over the available task information and written context. The LLM prompt contains the JSON-serialized narrative view from the previous step, along with a structured system prompt that specifies the report format. We define two prompt variants
\begin{itemize}
    \item Full realization prompt: An exhaustive structured report covering all known information (player status, active and completed patient needs, all entity and item locations, and any unresolved directional facts). This variant is designed to maximize information completeness.
    \item Task-aware prompt: A compressed report that foregrounds only outstanding task requirements, entity locations relevant to those needs, and directional facts, while omitting completed actions. This variant is designed to maximize task-relevant information density.
\end{itemize}

Both prompts enforce a rigid bulleted template to facilitate consistent downstream evaluation (i.e., ensuring that generated reports are amenable to the same structured extraction used for human reports).

\subsection{Experimental conditions}

\bh{To answer our research questions and isolate the contribution of each pipeline component, we evaluate five report conditions, outlined in Table \ref{tab:experiment_conditions}, decomposed by source input, intermediate representation, reconciliation, and renderer. The user report condition uses the outgoing participant's original free-text report as-is, serving as the unaided human baseline. The task-aware and full-realization conditions run the full reconciliation pipeline and differ only in the rendering prompt; their contrast addresses RQ3. The end-to-end ablation bypasses graph construction, alignment, reconciliation, and narrative view entirely, providing the raw system trace and participant report directly to the same LLM with equivalent task-aware prompting; its contrast with the task-aware condition holds the source inputs and generation instructions fixed and isolates the value of explicit structured reconciliation (RQ2). The telemetry-only ablation omits the participant's written report, running the full pipeline on telemetry alone; together with the user-report baseline it addresses source complementarity (RQ1).}

\begin{table}[]
    \centering
    \bh{%
    \begin{tabular}{l|c|c|c|c|l}
        Condition   & Human report & Telemetry & Structured repr. & Reconciliation & Renderer \\
        \hline
        User report (baseline) & x & -  & - & - & Human (verbatim) \\
        Task-aware (TA)        & x & x  & x & x & Task-aware LLM prompt \\
        Full-realization (FR)  & x & x  & x & x & Exhaustive LLM prompt \\
        End-to-end (end2end)   & x & x  & - & - & Task-aware LLM prompt \\
        Telemetry-only (TA -user) & - & x & x & x & Task-aware LLM prompt
    \end{tabular}}
    \caption{\bh{Experimental conditions, decomposed by source input, intermediate representation, explicit state reconciliation, and renderer. All generated conditions are paired transformations of the same 13 participant task states.}}
    \label{tab:experiment_conditions}
\end{table}
\section{\texorpdfstring{\bh{Data and Controlled Task}}{Data and Controlled Task}}

\subsection{\texorpdfstring{\bh{A controlled multitask handover environment}}{A controlled multitask handover environment}}

\bh{In order to examine how users prioritize and report task-relevant information while controlling for prior experience level and system state observability, we designed a controlled spatial multitask environment (inspired by information-work challenges documented in healthcare handover, but making no claim to clinical realism) that instantiates the information problems characteristic of handover: incomplete task progress at a deadline, multiple concurrent obligations, spatially distributed resources with non-trivial search costs, interruption, partial system observability, information available only through human observation, opportunity for memory error, and a required transfer to a successor. It deliberately does not instantiate other properties of real handover settings: domain expertise, clinical risk, shared professional schemas, interactive verbal clarification between sender and recipient, or longitudinal shifts.} In \bh{this environment,} participants control a player character in a top-down 2D \bh{world}. The player can move around inside the map using arrow keys and interact with items or non-player characters (NPCs) using the spacebar. The player must complete tasks with several sequential steps by interacting with NPCs and items, and navigating through the game environment, with the goal to complete as many of these tasks as possible within the time limit.

The player was tasked with interacting with 5 ``patients," each in a different room. The tasks for each patient required several sequential steps, some of which needed to be repeated by each player. First, the player needed to fetch a specific item for each patient from one of two central storeroom locations. After receiving this item, the patient would request information from a named NPC (representing another medical care professional working with the patient). The player needed to locate the correct NPC and interact with them, then bring the information back to the patient. Some patients then required information from a different NPC, or gave instructions for a step the next player needed to complete. Any tasks that were not completed by the first player would be handed off to the second. The game was designed so it was not possible for the player to finish all possible tasks, so there were always outstanding tasks (one per patient) that needed to be handed off to a potential next player. 

The map layout was loosely based on a hospital floor and the movement patterns required of nurses during a typical shift \cite{potter_analysis_2005}, with patient rooms, storerooms for items needed for patients, and lounges where non-patient characters were located. Because it took a non-trivial amount of time to travel down hallways from one room to another, completing the tasks efficiently required clarity of which tasks needed to be completed next and where they were located. Task information was structured in such a way that the player needed to keep track of at least one outstanding need per patient at any given time.

\begin{figure}
    \centering
    \includegraphics[width=1.0\linewidth]{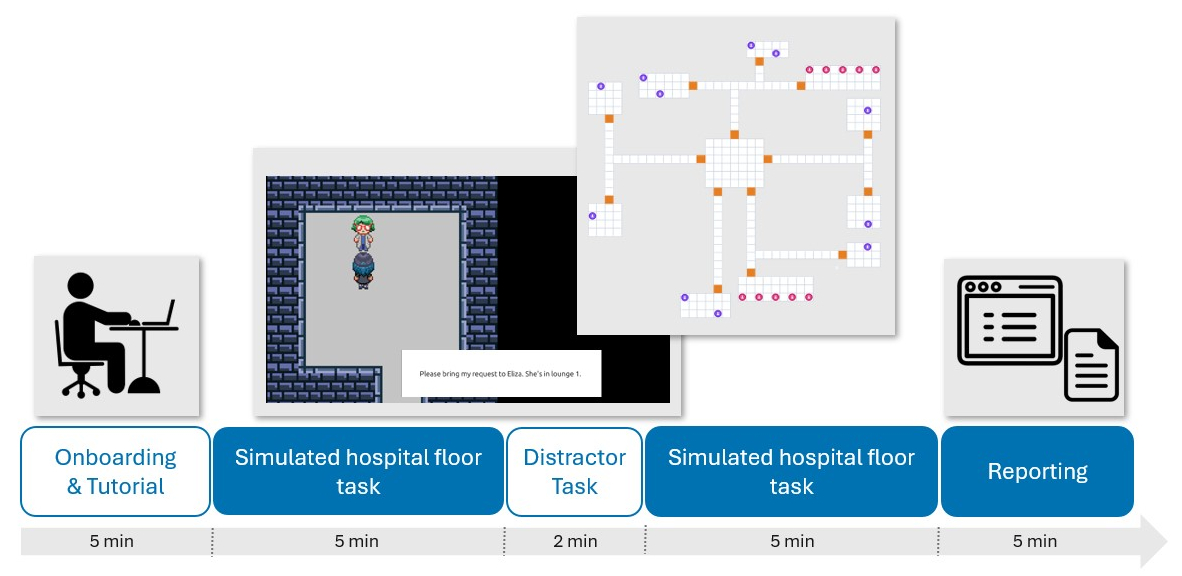}
    \caption{\kb{Walk-through of the data collection, showing the task interface and map. Participants controlled the player character with keyboard controls detailed in a short tutorial. Left: the interface includes a display of the current room, the player character (blue), and non-player characters within, along with the room name, their score (number of tasks complete), and their time progress in the task. Right: tasks required participants to navigate a map consisting of several rooms, interacting with non-player characters (indicated in purple) and item resources (indicated in pink). Participants did not have access to the map during the task. The environment creates concurrent, spatially distributed task threads and a controlled information asymmetry between system telemetry and human observation (Table \ref{tab:observability}). The task was also interrupted with a distractor task designed to further challenge participants' working memory of their task threads and progress. After 10 minutes of in-game time had elapsed, participants were asked to write a handover report detailing their task state and progress to a hypothetical successor.}}
    \Description{A figure showing the flow of the user data collection. There are 5 components with accompanying illusations: Onboarding and tutorial (5 minutes), Simulated hospital floor task, part 1 (5 minutes), Distractor task (2 minutes), Simulated hospital floor task, part 2 (5 minutes), and Reporting (5 minutes). A screenshot of the task user interface and an illustration of the map used for the game task are included.}
    \label{fig:game_task}
\end{figure}

\subsection{User data collection}

To evaluate our system and with approval from our institution's Institutional Review Board, we gathered sample task observation sets and handover reports from N = 13 participants recruited from the student population of our university. Participants first played a tutorial round of the game to familiarize them with the controls, which took approximately 1-3 minutes. When they finished the tutorial they were given 10 minutes to make as much progress in the game as possible. Halfway through their time, they performed a memory-matching distractor task for an additional 2 minutes. After 10 minutes of play, they were interrupted by a notification that their time had ended. At this point the task state was saved and they could not continue. 

Participants were then given 5 minutes to write a report that would allow another person, or themselves in the future, to complete the game from where they left off. They were told to assume the person continuing the task had no prior knowledge about the game environment. Otherwise they were not instructed what the report should contain or its format and style. Note that no matter how much progress someone had made in the game, 5 incomplete tasks needed to be handed off to the next participants (though participants might also include information about the environment layout, and their prediction of future tasks based on their experience).

The system trace from each user session was also recorded. This trace consisted of timestamped plain-text telemetry logging events like room transitions, initiation of NPC interactions, and item pickups and deliveries. Visual information (such as the positions or appearances of NPCs or items) and dialogue content (where player instructions, including NPC requests and directional hints, were communicated) were not logged and were thus only visible if included in user-written reports.

\bh{Each participant session yields one independent task state. All generated report conditions are paired transformations of these 13 states; the resulting 65 reports are therefore not 65 independent participant observations, and our sample supports controlled within-state comparison of report-generation strategies rather than population-level behavioral claims.}

\subsection{\texorpdfstring{\bh{Observability design}}{Observability design}}

\bh{Table \ref{tab:observability} summarizes which information each source can, in principle, capture in our environment. This asymmetry is a deliberate design choice: it creates a controlled test of source complementarity in which neither telemetry nor the human report is sufficient on its own. We note the corresponding interpretive boundary explicitly: because dialogue content was engineered to be invisible to telemetry, the benefit of fusing human input observed in our results may be larger than in domains where instrumentation is more complete (see Limitations).}

\begin{table}
    \centering
    \bh{%
    \begin{tabular}{l|c|c}
        Information type & Telemetry can observe & Human can observe/report \\
        \hline
        Room transitions & Yes & Yes \\
        Item pickup/delivery & Yes & Yes \\
        Dialogue content (requests, hints) & No & Yes \\
        Outstanding need inferred from dialogue & Partial & Yes \\
        User intent and strategy & No & Yes \\
        Exact event timestamps & Yes & Usually no \\
        Completed action history & Yes & Partial (memory-limited) \\
        Memory error & Not applicable & Possible \\
    \end{tabular}}
    \caption{\bh{Observability matrix for the controlled environment: which information types are available to system telemetry versus to the outgoing human. The engineered asymmetry makes the two sources complementary by design.}}
    \label{tab:observability}
\end{table}

\section{\texorpdfstring{\bh{Artifact-Level Evaluation of Estimated Task-Resumption Utility}}{Artifact-Level Evaluation of Estimated Task-Resumption Utility}}

\subsection{\texorpdfstring{\bh{Evaluation scope and estimand}}{Evaluation scope and estimand}}

\bh{We state the boundary of our evaluation up front: we evaluate whether a report \emph{contains task-state information expected to reduce reconstruction effort}; we do not observe a receiver. Concretely, our metrics estimate the task-state information available to a hypothetical recipient under an explicit search-cost model derived from the controlled environment's topology and task rules. The resulting quantities are artifact-level proxies for task-resumption readiness (not measurements of recipient behavior, time to resumption, error rate, or task completion), and we use ``task-resumption potential'' only in this estimated, model-relative sense. This artifact-level scope is chosen deliberately: it is what allows fully objective scoring, since every claim in a report is verified against a known ground-truth state rather than against rater judgment. Downstream recipient performance is the motivating application, not the measured outcome.}

\subsection{\texorpdfstring{\bh{Estimating task-state utility via information access cost}}{Estimating task-state utility via information access cost}}

The constrained nature of the simulated task allowed us to directly assess the value of state-related information that could be reconstructed from the reports across conditions. This was done by repeating the same fact-extraction process used in the generation pipeline (extracting facts into a domain-specific language and then converting to a semantic graph) on all the reports. DSL conversion was performed by LLM, with human co-annotation on 30 percent to verify annotation reliability (Cohen's $\kappa$ = .72). \bh{We considered the structural risk that an LLM-based evaluator could favor machine-generated reports, whose regular formatting may be easier to parse than free-form prose, and designed the evaluation to contain it. Because LLM judges are subject to known biases \cite{chen_humans_2024}, the model's evaluative role is confined to surface-form fact extraction; all scoring of extracted facts is performed deterministically against ground-truth state, so no model judgment enters the utility or misinformation calculations. Extraction itself is validated against human co-annotation, following recommended practice for justifying LLM-assisted annotation \cite{calderon_alternative_2025}, and identical extraction and scoring machinery is applied to every condition. Finally, our central contrasts (RQ2 and RQ3) compare machine-generated conditions that share the same rigid output template, so differential parseability cannot account for those results; only the comparison against free-form user reports inherits any residual extraction asymmetry, and the margins observed there substantially exceed what extraction error could plausibly produce given the validated agreement.}
We then analyzed the content of these reconstructions for their informational value, quantified in terms of relative Information Access Cost (IAC), a key factor in real-world handover preparation behavior \cite{yang_effect_2013}.

\bh{IAC is a task-grounded evaluation framework that estimates the operational value of a handover report by quantifying how much search effort a downstream agent would be spared, or would waste, if it acted on the reported knowledge graph $\mathcal{G}_{pred}$ in our spatial, multi-task environment. Unlike precision/recall metrics that treat all facts uniformly, IAC weights each fact by the expected search cost (including both physical movement and interaction time) an agent would face in the absence of that information, producing a score denominated in task-derived \emph{search-cost units}. Because these units arise partly from search-space cardinality and fixed interaction constants rather than literal simulated walking distances, we refer to them as cost units rather than ``steps'' throughout.}

\bh{We report two constructs separately, and deliberately name both as estimates:}
\begin{enumerate}
    \item \bh{\textbf{Estimated task-state utility} (cost saved, $C_{saved}$): the reduction in modeled reconstruction cost attributable to information in the report, i.e., cost that would otherwise be incurred by a recipient with no prior information. Its complement, $C_{max} - C_{saved}$, is the \textbf{estimated residual reconstruction cost} left to the recipient.}
    \item \bh{\textbf{Misinformation burden} ($C_{misinfo}$): the modeled penalty incurred by an agent acting on incorrect information, which can force wasted actions or searches of wrong areas.}
\end{enumerate}

$C_{saved}$ is \bh{assessed} on a per-patient basis, while $C_{misinfo}$ is assessed globally.


\subsubsection{Component types}
For each target patient $e \in \mathcal{E}$, the required information is decomposed into three orthogonal components:

\begin{enumerate}
    \item Location ($C_{loc}$): Where is the patient located?
    \item Need ($C_{need}$): What specific medication or message-routing task does the patient require?
    \item Resource ($C_{res}$): Where is the resource (e.g., medication or delivery target) needed to resolve the patient's need?
\end{enumerate}

Each component is scored independently by comparing the predicted facts in $\mathcal{G}_{pred}$ against the ground-truth game state $\mathcal{S}_{gt}$ and the environment topology $\mathcal{T}_{map}$.

\subsubsection{Baseline information costs}
For spatial information ($C_{loc}$, $C_{res}$), baseline search costs represent the expected \bh{search cost (in search-cost units)} of searching the environment for a target category, depending on the target entity type (i.e. searching all patient rooms for a patient, or all storage rooms for a medication).
If the entity type is unknown, $C_{max} = \frac{|\mathcal{R}_{all}|}{2.0}$, where $\mathcal{R}_{all}$ is the set of all rooms in the environment.

For diagnosis information ($C_{need}$): A fixed baseline cost of $30.0$ representing the diagnosis interaction time. This constant was derived from the expected steps needed to navigate to a known patient's room and complete the dialogue interaction needed to find the patient's needs. \bh{Like all parameters of the cost model, it is an explicit, fixed modeling choice applied identically to every condition and report, so paired comparisons between conditions are made under the same assumptions.}

\subsubsection{Formulation \& Scoring Matrix}
Let $C_{max}(e, c)$ be the baseline search cost for a given entity $e$ and component $c$ if the agent had no report and had to search the environment blindly.

\begin{table}
    \begin{tabular}{l|l|l}
        Credit Type ($\kappa$) &	Condition  & $C_{saved}$	\\ 
        \hline
        FULL	& Perfect match or correct omission	& $C_{max}$	\\
        PARTIAL	& Correct but vague or under-specified information	& $P \cdot C_{max}$	\\
        NONE	& No correct relevant information provided	& $0$	\\
    \end{tabular}
    \caption{Cost breakdown by classification of information extracted from handover reports.}
    \label{tab:cost_breakdown}
\end{table}

For each component, the predicted facts are mapped to a Credit Type $\kappa \in \{\text{FULL}, \text{PARTIAL}, \text{NONE}\}$, and a partial credit factor $P \in [0,1]$, which determine $C_{saved}$, the portion of baseline search effort the recipient agent avoids thanks to the reported information, per the schema in Table \ref{tab:cost_breakdown}. \bh{We define a \emph{correct omission} as referring to a component for which the ground-truth state contains no outstanding information to convey (for example, a need that the outgoing player already resolved): a report that says nothing about such a component has left nothing for the recipient to rediscover, so the component's full baseline value is credited. Facts about entities or components outside this task-defined target universe are not scored as omissions; they are simply out of scope for $C_{saved}$ (though they remain subject to the misinformation check below).}

\subsubsection{Location Scoring ($C_{loc}$ and $C_{res}$)}
Both the patient NPC location and the resource location components use identical spatial verification logic. When exact location information was not provided in handover reports, users frequently employed referring expressions that could be used to completely or partially constrain the search space for the recipient, a typical occurrence in natural language communication \cite{campana_natural_2011}. Both definite and relative location information was evaluated as follows.

\paragraph{Fact selection.} All facts in $\mathcal{G}_{pred}$ that reference NPC $e$ are collected, including direct room assignments (e.g., "Lily is in Room 1"), directional constraints (e.g., "Lily is north of the hallway"), and embedded location constraints on subject/target arguments (e.g. "someone in room 1 who needs..."). When multiple facts reference the same NPC, the system selects the single most specific fact.

\paragraph{Entity resolution.} Arguments in facts may be named (e.g., "lily") or existential (e.g., "a patient in room 1"). An existential argument resolves to NPC $e$ if (1) any type-filtering heuristic is satisfied (medication names match items; person references match NPCs), and (2) either no spatial constraint is attached (unconstrained existential), or $e$'s true room satisfies the spatial constraint given $\mathcal{T}_{map}$.

\paragraph{Spatial constraint verification and score assignment. } Let $constraint(f)$ be the spatial condition asserted by the selected fact $f$. The set of rooms satisfying this constraint in the map graph is defined as $\mathcal{R}_{sat} \subseteq \mathcal{R}_{all}$. Let $\mathcal{R}_{baseline}$ be the canonical search rooms for that entity type, and let $r_{true}$ be the true room of the entity in $\mathcal{S}_{gt}$.

$$\kappa = \begin{cases} 
    \text{FULL} & \text{if } r_{true} \in \mathcal{R}_{sat} \text{ and } |\mathcal{R}{sat} \cap \mathcal{R}_{base}| = 1 \\ 
    \text{PARTIAL} & \text{if } r_{true} \in \mathcal{R}_{sat} \text{ and } |\mathcal{R}{sat} \cap \mathcal{R}_{base}| > 1 \\ 
    \text{NONE} & \text{if no correct matching fact exists} \end{cases}$$

For partial matches, the credit factor is computed as the search-space reduction ratio:

$$P = 1 - \frac{|\mathcal{R}_{sat} \cap \mathcal{R}_{base}|}{|\mathcal{R}_{base}|}$$

This rewards facts that narrow the search space without pinpointing a unique room.

\subsubsection{Entity Need Score ($C_{need}$)}
The system evaluates whether the report accurately details the outstanding need (e.g., patient needs a specific medication, has a message to be delivered elsewhere, or is awaiting a response). Because both human- and LLM-written reports often use pronouns or generic descriptions, the pipeline maps predicted arguments to the target entity $e$ using a three-tier strength model and scored accordingly:

\begin{enumerate}
    \item Definite: Direct named match (e.g., "lily" matches Lily), or a spatially constrained existential argument (e.g., "someone in room 1") resolving exclusively to $e$.
    \begin{itemize}
        \item FULL: Predicate matches (e.g., \textit{needs-medicine} or \textit{has-message-for}) and the associated resource/partner is an exact match.
        \item PARTIAL: Predicate matches, but a generic category is used instead of the specific resource (e.g. "medicine" instead of "gold medicine"), earning a static $P = 0.5$ credit.
    \end{itemize}
    \item Possible: An unconstrained existential (e.g., "someone") or a constrained existential resolving to multiple candidates including $e$.
    \begin{itemize}
        \item Capped at PARTIAL credit ($P = 0.5$) only when the resource is an exact match. It cannot trigger a contradiction penalty, which prevents general statements (e.g., "someone needs the red medicine") from penalizing other patients.
    \end{itemize}
    \item None: Explicitly matches a different entity, or an existential that excludes $e$.
\end{enumerate}

\subsubsection{Additional misinformation penalty}
While the scoring algorithm thus far has focused on strictly necessary facts for task continuation, inaccurate information (such as invented patients or inaccurate supplementary details) can also incur a significant burden on a handover recipient's ability to efficiently continue the task. For that reason, in addition to the per-entity component scores, a global misinformation cost $M_{global}$ is computed by independently verifying every predicted fact in $\mathcal{G}_{pred}$ against $\mathcal{S}_{gt}$. For spatial and location facts, the system checks whether any NPC or item matching the argument (by name or by existential type+constraint) actually occupies a room satisfying the stated constraint. If no match exists, the fact is flagged as misinformation. For facts relating to outstanding patient needs, the system checks whether the asserted relation matches any patient NPC's current ground-truth need. If no match is found, the fact is flagged.

Each flagged fact incurs a cost equal to the baseline search cost for its category:

$$M_{global} = \sum_{f \in \mathcal{F}_{misinfo}} C_{baseline}(\text{type}(f))$$

where $F_{misinfo}$ is the set of all flagged facts and $C_{baseline}$ is the NPC/item-type-specific search cost for location facts, or the constant diagnosis cost for need facts.

\subsubsection{Metric Aggregation}
The total cost saved is summed across all patient NPCs and their components:

$$S_{total} = \sum_{e \in \mathcal{E}} \sum_{c \in {loc, need, res}} C_{saved}(e, c)$$

The theoretical maximum (i.e., the total cost saved by a perfect report) is the sum of all baseline search costs:

$$C_{max}^{total} = \sum_{e \in \mathcal{E}}   \sum_{c \in {loc, need, res}} C_{max}(e, c)$$

which is a fixed constant for a given environment configuration and degree of task progress. \bh{Because task progress differs across sessions, $C_{max}^{total}$ can differ across participants; a normalized estimated task-state utility can accordingly be defined per task state as $TSU = S_{total} / C_{max}^{total}$. Within a given task state the denominator is constant across conditions, so the within-state ordering of conditions, the basis of our paired analyses, is unaffected by this normalization.}

For the purposes of summarization, we also define an aggregate \textbf{combined cost} to express the total operational penalty an agent combining the residual search effort (what the report failed to eliminate) with the misinformation penalty:

$$C_{comb} = \left(C_{max}^{total} - S_{total}\right) + \alpha \cdot M_{global}$$

The first term is the unsaved baseline cost, the search effort that remains because the report was incomplete or incorrect. The second term scales the global misinformation cost by $\alpha = 3$, reflecting that acting on false information in this task domain is approximately three times as costly as lacking information entirely (the agent must travel to the wrong location and then resume an uninformed search). \bh{The value of $\alpha$ is a manually selected modeling choice; we therefore treat the combined index as descriptive only and draw all inferential conclusions from the separately reported components ($S_{total}$ and $M_{global}$), which makes those conclusions independent of the choice of $\alpha$ by construction.}

\subsection{\texorpdfstring{\bh{Task-value density and report length}}{Task-value density and report length}}
\bh{To address RQ3, we assessed the task-value density of reports: the estimated task-state utility conveyed per token of output,}
$$C_{\bh{density}} = \frac{C_{saved}}{tokens}$$

\bh{Token counts were calculated with the \texttt{o200k\_base} tokenizer used by the GPT-4.1 model family. We emphasize that this is a density measure over task-derived cost units, not information-theoretic compression, and that a ratio alone can reward degenerately short outputs; we therefore interpret it only alongside the absolute utility and misinformation results, and report raw report lengths with the results.}

\subsection{Characterizing information content of user reports}
\bh{To address our exploratory question about what human reports contribute beyond task state,} user-generated reports were annotated using a simple schema to assign a category label to each clause according to the nature of information contained within:

\begin{itemize}
    \item S - State Transfer: A factual, falsifiable claim about the current world state at handover time.
    \item K - Knowledge Transfer: Something the user learned or believes from experience that a fresh agent with full state visibility would not automatically know. Includes strategy, priorities, causal knowledge ("you need X to do Y"), dead ends already ruled out, warnings, and predictions.
    \item M - Meta/Other: About the report or the user's experience rather than the task. Includes confidence hedges, apologies, comments about the task framework, and filler.
    \item A - Ambiguous/Mixed: Has features of multiple of the above with unclear distinction.
\end{itemize}

Annotations were performed via LLM (GPT 4.1-mini) with human co-annotation on 30 percent of reports to ensure reliability (Cohen's $\kappa$ = .64 overall, .84/.74 on state and knowledge categorization). \bh{Consistent with recommended practice for validating LLM-assisted annotation against human annotators \cite{chen_humans_2024, calderon_alternative_2025}, we report agreement for the categories on which our observations rest; this analysis is exploratory and descriptive, and none of the paper's primary claims depend on it.}

\subsection{\texorpdfstring{\bh{LLM configuration and reproducibility}}{LLM configuration and reproducibility}}
\bh{Four pipeline or evaluation stages invoke an LLM: (1) normalization of user reports into the DSL, (2) existential-reference resolution during entity alignment, (3) report generation, and (4) evaluation-time fact extraction and content annotation (GPT-4.1-mini for annotation, as noted above). All other transformations are deterministic given the LLM outputs. Prompts, few-shot examples, output schemas, alias tables, and parsing rules for each stage are released as supplementary material\bh{, consistent with calls for transparent, multi-metric evaluation of LLM-based systems \cite{liang_holistic_2022}}. } \kb{LLM calls were made via the OpenAI API ChatCompletions endpoint "gpt-4.1-mini" with knowledge cutoff of 01-Jun-2024, accessed 18-Jul-2026, with 0 temperature; outputs were processed as-is with no retry policy or manual repair of outputs.}
\section{Results}

\bh{We organize the results by research question. Friedman tests were used to evaluate omnibus differences across the five report conditions for each outcome, with post-hoc pairwise analysis performed using Wilcoxon signed-rank tests with Holm correction applied within each outcome's family of pairwise comparisons. Omnibus differences were significant for estimated cost saved, $\chi^2(4, N=13) = 33.60$, $p < .001$, Kendall's W $= 0.65$, for weighted misinformation cost, $\chi^2(4, N=13) = 31.95$, $p < .001$, Kendall's W $= 0.61$, and for task-value density, $\chi^2(4, N=13) = 29.46$, $p < .001$, Kendall's W $= 0.57$. Because our combined cost index depends on the manually selected misinformation weighting ($\alpha$), we report it for descriptive purposes only (Table \ref{tab:combined-cost}) and reserve statistical analysis for the component-level outcomes.} \label{cost-reporting}

\begin{table*}[t]
\centering
\caption{Descriptive cost metrics by condition. Combined cost is computed at $\alpha = 3$ and reported for descriptive purposes only; \bh{estimated} cost saved and misinformation cost are reported separately as primary inferential metrics (see \ref{cost-reporting}). Values are median [range] across $n=13$ participants. \bh{Units are task-derived search-cost units, not literal movement steps.}}
\label{tab:combined-cost}
\small
\begin{tabular}{@{}lccc@{}}
\toprule
Condition & Combined cost \bh{(search-cost units)} & Omission share (\%) & Misinformation share (\%) \\
\midrule
Baseline (user-only)       & 1150.51 \; [22.50, 2023.93]  & 64.6 \; [34.7, 100.0] & 35.4 \; [0.0, 65.3] \\
\textbf{Task-aware}        & 370.21 \; [150.00, 965.94]  & 63.0 \; [17.6, 100.0] & 37.0 \; [0.0, 82.4] \\
Telemetry-only             & 274.70 \; [150.00, 551.50]  & 100.0 \; [34.7, 100.0] & 0.0 \; [0.0, 65.3] \\
Full-realization           & 361.20 \; [191.50, 789.24]  & 44.6 \; [18.2, 100.0] & 55.4 \; [0.0, 81.8] \\
End-to-end                 & 773.20 \; [370.21, 1345.51] & 32.5 \; [14.2, 50.4]  & 67.5 \; [49.6, 85.8] \\
\bottomrule
\end{tabular}
\end{table*}

\begin{figure}
    \centering
    \includegraphics[width=0.85\linewidth]{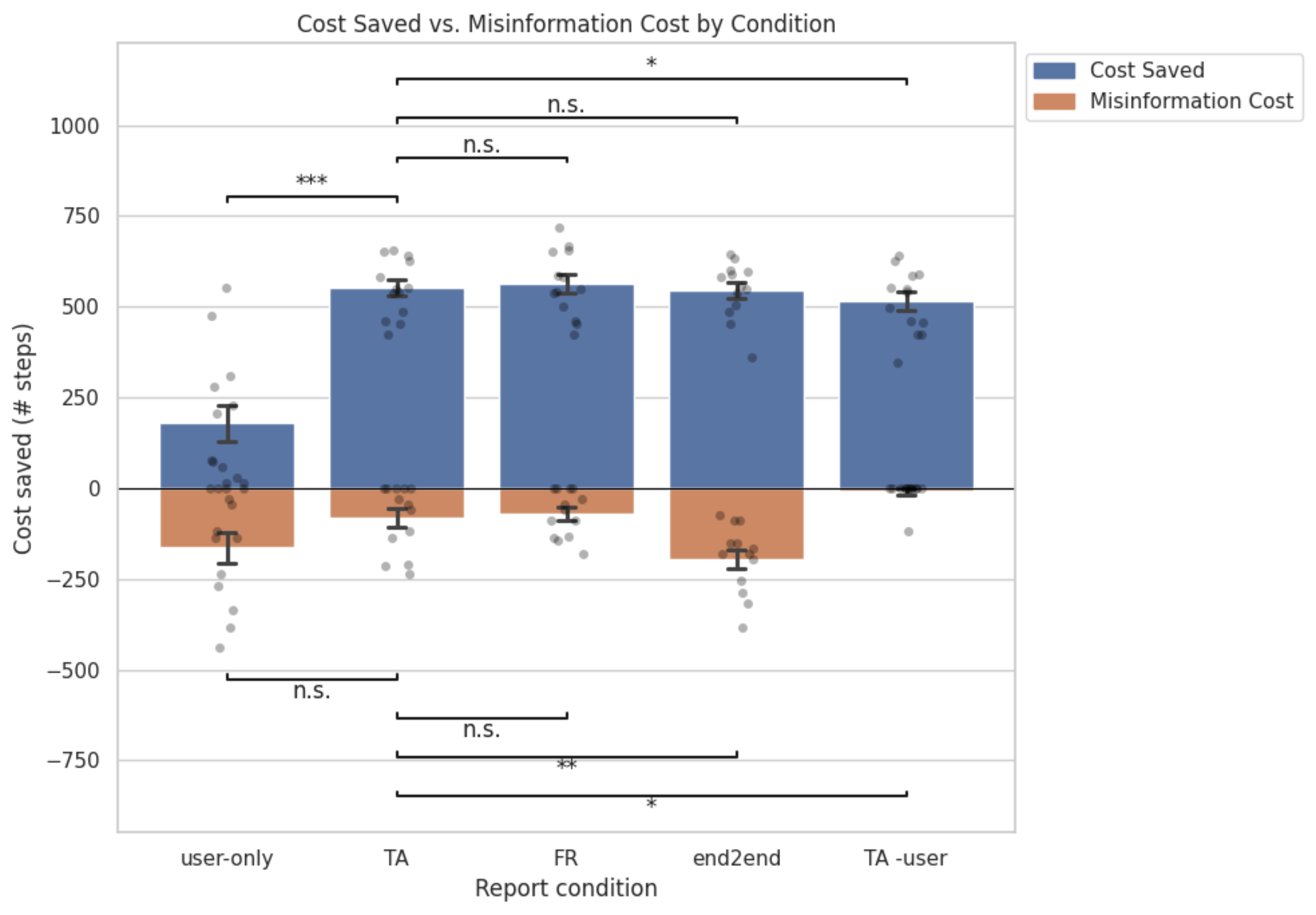}
    \caption{\bh{Estimated information access cost saved (positive axis) and misinformation cost incurred (negative axis) by report condition. TA = task-aware (ours); end2end = end-to-end (LLM-only ablation); TA -user = telemetry-only ablation; n.s. = not significant; * = $p < .05$; ** = $p < .01$; *** = $p < .001$. Task-aware reports saved significantly more estimated cost than the user-only baseline and the telemetry-only ablation, and incurred significantly less misinformation than the user-only baseline and the end-to-end ablation. Estimated cost saved did not differ significantly between the task-aware, full-realization, and end-to-end conditions, and the telemetry-only ablation incurred the least misinformation of all conditions.}}
    \Description{A bar chart with one bar group per report condition. Bars above the horizontal axis show estimated cost saved and bars below the axis show misinformation cost incurred, with significance brackets over selected condition pairs.}
    \label{fig:cost_saved_vs_misinfo}
\end{figure}

\subsection{\texorpdfstring{\bh{RQ1: Source complementarity}}{RQ1: Source complementarity}}

\bh{Reconciling the two sources preserved more estimated task-state utility than either source alone. Estimated cost saved was significantly higher for reports generated by the fused task-aware pipeline (\textit{Mdn} = 548.64 cost units) than for participants' own reports (\textit{Mdn} = 158.87, $p_{adj} < .001$, $r = 1.0$), and significantly higher than for reports generated from telemetry alone ($p_{adj} = 0.01$, $r = 1.0$). Under this task and metric, the two sources therefore contributed complementary state information: neither the human report nor the telemetry alone recovered the estimated utility available from their reconciliation. We emphasize that this is a claim about the state information recoverable from the report artifacts under our cost model, not a claim about recipient performance.}

\bh{Complementarity came with a measurable tradeoff.} Total weighted misinformation costs differed significantly across report types\bh{, and the ablation that omitted human input entirely (telemetry-only) incurred the lowest misinformation cost of any condition (\textit{Mdn} = 0.0); task-aware reports, which incorporate the user report, incurred significantly more (\textit{Mdn} = 45.67, $p_{adj} = .039$, $r = 1.0$).} Because trace-only reports consistently contained little to no misinformation, most of the incorrect information in the task-aware condition was \bh{inherited from} the user reports, which frequently contained misremembered names, locations, and task details. While the \bh{state reconciliation} pipeline was able to address some of these mistakes, the task as designed contained some information that was only available to the user and not observable from system telemetry. As a result, the task-aware pipeline could not verify the correctness of some user input, and would preserve this incorrect information in \bh{its} output. \bh{Reconciliation nevertheless reduced the inherited burden substantially relative to the unaided baseline: task-aware reports accrued significantly lower misinformation cost (\textit{Mdn} = 45.67) than participants' own reports (\textit{Mdn} = 135.67, $p_{adj} < .001$, $r = .86$), reflecting cases where telemetry could contradict misremembered content.} \bh{Adding human-only information thus increased estimated utility but also introduced unverifiable error, a value--risk tradeoff rather than a strict improvement, which we return to in the Discussion.}

\subsection{\texorpdfstring{\bh{RQ2: Structured reconciliation versus direct synthesis}}{RQ2: Structured reconciliation versus direct synthesis}}

\bh{Given identical source inputs and equivalent task-aware prompting, the structured pipeline and the end-to-end LLM produced reports with statistically comparable estimated cost saved ($p_{adj} = .33$). The two conditions differed sharply on misinformation: task-aware reports accrued significantly lower weighted misinformation cost (\textit{Mdn} = 45.67) than end-to-end reports (\textit{Mdn} = 180.0, $p_{adj} = .004$, $r = .93$); the structured condition's median was roughly one quarter of the end-to-end median. Structured reports were also significantly denser in estimated utility per token (\textit{Mdn} = 2.64 vs. 2.30, $p_{adj} = .01$, $r = .87$). In other words, on these data explicit reconciliation did not significantly change how much estimated task-state utility the model could produce from the same inputs; it changed the \emph{failure profile} of the synthesis.}

The end-to-end ablation, which had access to the same input data but contained no explicit reconstruction, reconciliation, or consolidation of model state, \bh{preserved} mistakes introduced by the user while also generating many new hallucinated facts \bh{absent from both inputs}. This behavior indicates the value of explicit identification and reconciliation of contradictory facts, as well as the importance of providing structured input to avoid hallucinations. \bh{The misinformation-source observations reported here (user-inherited versus generation-added) are based on manual inspection of the facts flagged in each condition.}

\subsection{\texorpdfstring{\bh{RQ3: Task-aware versus exhaustive realization}}{RQ3: Task-aware versus exhaustive realization}}

\bh{As a manipulation check} that the differing prompt types did result in differing \bh{content selection} (rather than simply condensing a similar fact set), we calculated precision/recall on extractable state facts against those included in the LLM prompt.
A Wilcoxon signed-rank test indicated that recall was significantly lower for reports generated by the task-aware prompt (\textit{Mdn} = 0.52) set compared to those generated by the full-realization prompt set (\textit{Mdn} = 0.68, $W = 1.0$, $p_{adj} < .001$, $r_{rb} = .96$). Precision did not differ significantly ($p = .98$). \bh{These results confirm that the task-aware prompt selectively omitted available facts as instructed, without introducing additional unsupported content. We note that recall against the prompt's own fact set is a check on the rendering manipulation, not a measure of ground-truth handover quality; lower recall here reflects intended selectivity, not a deficiency.}

\bh{Despite including fewer of the available facts, task-aware reports preserved estimated utility that was statistically comparable to the much longer full-realization reports ($p_{adj} = .96$), and conveyed that utility more efficiently: estimated cost saved per token was significantly higher for the task-aware prompt (\textit{Mdn} = 2.64) than for the full-realization prompt (\textit{Mdn} = 1.56, $p_{adj} < .001$, $r = 1.0$). Task-aware density did not significantly exceed the telemetry-only condition (\textit{Mdn} = 2.77, $p_{adj} = .52$), as expected given that telemetry-only reports draw on strictly less source information and are correspondingly short. Task-aware rendering therefore achieved a better utility--length tradeoff than exhaustive rendering of the same reconciled state; whether such filtering benefits an actual recipient's comprehension remains untested.}

\subsection{\kb{Comparison with frontier reasoning model}}

\begin{table}[htbp]
\centering
\caption{Comparison with frontier model, combined cost and omission/misinformation shares by condition}
\label{tab:frontier-cost-comparison}
\begin{tabular}{lccc}
\toprule
Condition & Combined Cost & Omission Share (\%) & Misinformation Share (\%) \\
\midrule
Task-aware (GPT 4.1-mini)       & 370.21  [150.00, 965.94]  & 63.0  [17.6, 100.0] & 37.0  [0.0, 82.4] \\
Task-aware (GPT 5.6-terra)        & 233.20 [101.50, 485.90]  & 100.0 [36.8, 100.0]  & 0.0 [0.0, 63.2]   \\
\textbf{End-to-end (GPT 5.6-terra) }  & 203.20 [116.50, 444.40]  & 100.0 [36.5, 100.0]  & 0.0 [0.0, 63.5]   \\
\bottomrule
\end{tabular}
\end{table}

\kb{In order to compare model behavior and the performance profile of the pipeline on a frontier reasoning model, we re-generated reports for the task-aware and end-to-end conditions using GPT 5.6-terra (accessed via OpenAI's Responses API endpoint "gpt-5.6-terra" 18-Jul-2026, with knowledge cutoff 16-Feb-2026 at time of writing) with medium reasoning. Evaluation DSL extraction was performed with GPT 4.1-mini for consistency. The frontier model performed better on cost saved and misinformation metrics on both conditions. Most notably, the end-to-end single-prompt approach (which resulted in increased misinformation with GPT 4.1-mini) now outperforms all other conditions on cost saved, with no increased misinformation. When examining the reports generated in this condition, the frontier model was consistently able to draw inferences across report and telemetry data, indicating that the benefit offered by the pipeline was sufficiently replicated by the reasoning capabilities of the frontier model (and vice versa).} 

\kb{To evaluate whether GPT 4.1-mini with our task-aware pipeline achieves performance comparable to the frontier end-to-end condition on estimated utility, we report the Hodges-Lehmann estimate of the paired difference alongside its 95\% CI, interpreted against a pre-specified margin of practical equivalence (25 percent of the Hodges-Lehmann estimate of the paired user-only vs. task-aware, GPT 4.1-mini difference on the corresponding metric), in the interest of anchoring the equivalence data to an effect size already established as meaningful within our own data. Given the modest sample size, we treat this as a descriptive comparison rather than a formal equivalence test (e.g., TOST), which would be underpowered at this n to support a confirmatory claim. } \kb{For cost saved, the margin was $\pm88.4$, with the calculated paired difference between frontier end-to-end and task-aware GPT 4.1-mini at -7.5 (95\% CI [-30.0, 7.5]), n=13 pairs, falling comfortably within the margin. For misinformation cost, the paired difference between the two was inconclusive due to the heavily zero-weighted distribution for both conditions (HL estimate 0.0, but with a large CI as 6 out of 13 participants had 0 misinformation in any reports).}

\kb{These results suggest that our pipeline architecture allows a smaller, non-reasoning model (GPT-4.1-mini) to approach the cost-saving performance of a larger, frontier-tier reasoning model operating end-to-end — though we treat this as a descriptive indication of comparability rather than a confirmed equivalence claim, given the sample size. These results position architectural structure as a partial substitute for model scale on this metric.}

\subsection{\texorpdfstring{\bh{Exploratory analysis: information content of user handover reports}}{Exploratory analysis: information content of user handover reports}}

\begin{figure}
    \centering
    \includegraphics[width=0.8\linewidth]{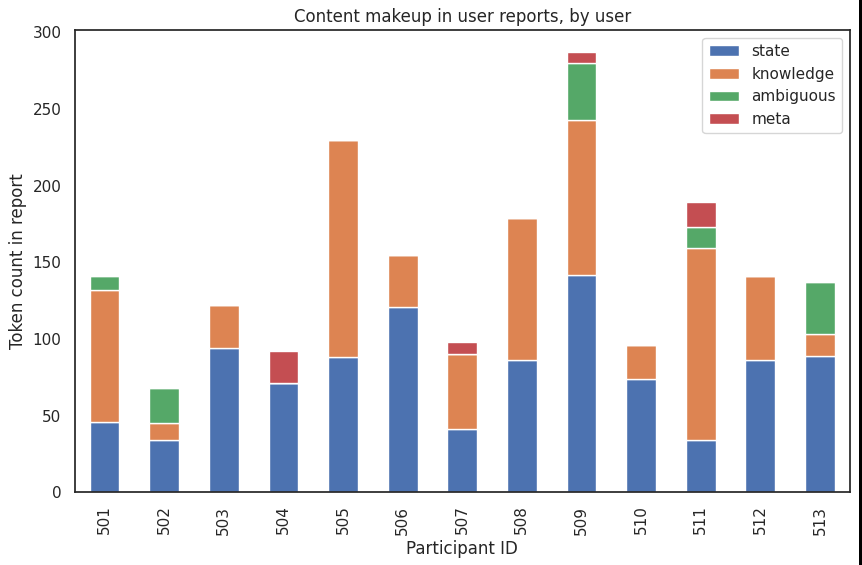}
    \caption{\bh{Breakdown of content types of user reports across all participants, by token count. Human-authored reports contained both task-state content and strategic-knowledge content. State was the largest category on average, while the proportion of strategic content varied substantially across participants. The value of this strategic knowledge is more difficult to evaluate objectively, and it is not captured by our state-focused pipeline or metrics.}}
    \Description{A stacked bar chart showing, for each participant's report, the proportion of tokens categorized as state transfer, knowledge transfer, ambiguous, or meta content. The mix varies widely across participants.}
    \label{fig:content_makeup}
\end{figure}

When faced with time pressure and the inherent ambiguity of the task, participants did not restrict themselves to solely conveying state information. \bh{Rather, many devoted a substantial portion of their report to strategic knowledge gained from their experience with the task.} On average, knowledge-transfer content made up 35 percent of each report's clauses (range: 0-66 percent), compared to 56 percent for state-transfer content (range: 18-78 percent). Content labeled as "ambiguous" or "meta" was minimal, averaging under 7 percent and 4 percent respectively. As indicated by the wide range in percent values, the split was not uniform across participants: several reports were dominated by a single category, while others showed a more even mix, suggesting individual variation in reporting strategy under time pressure rather than a single dominant style.

Content labeled as "knowledge-transfer" most often included recommendations for approaching the task (\textit{"Focus on one patient at a time, but remember all of the [medications] as you'll need them again later"}) as well as observations of tendencies in the map layout ("The patient rooms are all off the main room, but you'll often need to check waiting rooms") or patient needs ("Most patients need you to deliver a message to someone and then return with the response"). \bh{We interpret these data descriptively: state content was the largest category on average, while strategic knowledge constituted a substantial minority with strong individual variation. The data do not support a claim that participants reported strategy \emph{in place of} state; what they do show is that human reports contain a class of contribution (strategy, priorities, causal knowledge, and warnings) that our state-focused pipeline currently sidelines and our state-recovery metrics do not credit.} We return to this distinction in the Discussion, where we consider its implications for interpreting the precision/recall and cost results reported above.

\section{Discussion}

\subsection{\texorpdfstring{\bh{Structured state reconciliation as a reliability pattern}}{Structured state reconciliation as a reliability pattern}}
Task handover has long been understood as a problem of distributed situation awareness: the challenge of ensuring that an incoming agent has sufficient knowledge of the world state to continue a task effectively. Prior work has approached this challenge primarily through standardization, using structured forms, recall mnemonics, and training protocols that aim to make human reporting more complete and consistent \cite{clark_identified_2019}. These interventions treat handover failure as a problem of individual recall and communication discipline. \bh{Our results suggest a complementary framing: part of the difficulty of handover is a \emph{state reconciliation} problem, in which partial, differently reliable accounts of the same task must be integrated into a single defensible description of current state before anything is communicated at all.}

In our system, the goal of communicating a single, reconstructable model state is made explicit, as is the necessary prerequisite of integrating information from multiple sources into a single representation. Both the system trace and the user-written report consist of partial views of the task state; the former can be assumed to be accurate, but has only partial observability into the world (reflecting the limitations of \bh{analogous} real-world electronic records and information support systems), while the latter captures more nuanced information of intent and learned heuristics but is subject to memory error and omission, as we saw in the reports we collected. \bh{Neither source alone recovered the estimated task-state utility available from their reconciliation, and the fused output partially controlled, though it did not eliminate, the misinformation that human input introduced.}

\bh{Our comparison with the end-to-end ablation locates where this structure matters.} While it is superficially true that LLMs, especially when given access to system records, can improve the completeness of handover reports, our analysis demonstrates the value of using structured input and explicitly reconciling partial world models\bh{: given identical inputs and equivalent prompting, the structured pipeline matched the end-to-end condition's estimated state utility while incurring significantly less misinformation and conveying utility more densely. The benefit of structure on these data was thus primarily \emph{error control}, not coverage.} We attribute this to two properties of the pipeline that cannot be replicated by a single LLM call. First, by explicitly identifying conflicts between the user and system records, the pipeline surfaces contradictions as specific artifacts that the generation step can reason about directly. A single-call approach has no analogous mechanism; contradictory information simply arrives in the context window, and the model must either implicitly resolve or silently propagate the conflict. \bh{Our inspection of flagged facts suggests it frequently does the latter, preserving user-introduced errors while also generating new hallucinated facts not present in either input.} Secondly, the narrative view representation passed to the LLM is a consolidated and state-reconciled version of the input information, rather than raw telemetry logs and freeform text. This reduces the generation step to a well-scoped summarization task, limiting the surface area for hallucination. \bh{We do not claim that a knowledge graph is uniquely suited to this role; what matters are the properties the intermediate representation enforces (typed facts, provenance, temporal status, and inspectable conflicts), which other structured representations could provide equally well.}

These findings have practical implications for designers of AI-assisted handover tools, and more broadly for LLM-based systems that must synthesize information from multiple sources of varying reliability. The temptation to treat LLM generation as a general-purpose solution capable of handling input heterogeneity, conflict resolution, and summarization in a single pass may be counterproductive when input sources are known to conflict. Decomposing the task into explicit reconciliation and generation stages, even at the cost of increased pipeline complexity, appears to produce meaningfully \bh{safer} outputs. This is consistent with broader findings in the LLM literature on the benefits of structured intermediate representations for complex reasoning tasks \cite{delorenzo_abstractions--thought_2025}\bh{, and with work on grounded and attributed generation \cite{gao_enabling_2023, yuan_lcds_2025},} and suggests that handover is a domain where that decomposition is particularly valuable given the real-world costs of misinformation to the recipient.

\subsection{\texorpdfstring{\bh{Source complementarity is a value--risk tradeoff}}{Source complementarity is a value--risk tradeoff}}

\bh{Our results caution against reading either input source as simply ``better.'' Telemetry-only reports incurred essentially no misinformation but recovered less estimated utility, because parts of the task state were unobservable to the system by design. Incorporating the human report expanded what the output could cover (including content available in no log), but it was also the dominant source of the misinformation that survived reconciliation. Human input, in other words, expands observability while adding uncertainty. The system's role is therefore not to replace either source but to make their relationship explicit: corroborating what it can, flagging what it cannot, and carrying provenance forward so that downstream consumers can weigh unverified content appropriately.}

This framing has implications beyond the specific pipeline and test environment we investigate. Our results suggest that the design of AI-assisted handover tools (and information support tools more broadly) should foreground the relationship between human-provided and system-observed information, rather than treating LLM generation as post-hoc summarization over a single input. This is particularly true in real-world settings where electronic records may not be as accurate or up-to-date as human users. Where inputs conflict, the system's behavior (whether to flag, resolve, or silently prefer one source) becomes a consequential design choice with direct effects on recipient trust and task outcomes. Future systems might adopt a more human-in-the-loop approach and make this reconciliation visible to the outgoing agent, allowing them to identify and correct discrepancies before the report is finalized, rather than delegating resolution entirely to the model. \bh{Our current implementation preserves provenance and conflict records internally but does not render them for the reader; surfacing them (e.g., separating corroborated state from unverified human-only content and listing unresolved conflicts explicitly) is a natural extension that our representation already supports.}

\subsection{\texorpdfstring{\bh{State transfer, knowledge transfer, and the division of labor}}{State transfer, knowledge transfer, and the division of labor}}
While user-written reports frequently omitted or misremembered state information, resulting in low scores on our evaluation of actionable state data, our content analysis suggests that these reports were not predominantly empty or uninformative. Instead, user-written reports \bh{often devoted a substantial share of their content} to what we categorized as strategic knowledge transfer content: strategic priorities, learned heuristics, and general summaries that a receiving agent could still benefit from. Because our pipeline schema and evaluation focused so heavily on state reconstruction, this kind of abstracted information was not captured by either. \bh{This is a real cost of the current design: the system extracts state-related content and therefore discards or sidelines part of what humans contribute.} While this behavior was almost certainly influenced by the nature of our evaluation task - which, by design, participants had no prior experience with, leading them to highly value experiential guidance - it is also generally consistent with handover behavior in real-world contexts, where recommendations and strategic guidance are also considered valuable even among highly experienced professionals \cite{patterson_handoff_2004}.

This distinction in user behavior is useful for understanding the potential support role of handover support tools that focus on state reconciliation. \bh{The design implication we draw is a division of labor: the system reconstructs and verifies current task state, an activity where telemetry and cross-source checking outperform fallible recall; the human supplies intent, strategy, causal knowledge, and warnings, which no log contains; and the interface preserves human-only contributions verbatim while marking them as unverified rather than silently discarding or homogenizing them.} This also has implications for how such systems should be evaluated, as \textbf{metrics grounded purely in state fact recovery will systematically undervalue user contributions.} Future work should develop evaluation frameworks that treat the two content types separately, allowing for a fuller accounting of what human contributors bring to handover reporting. \bh{A concrete next step, achievable with our existing data, is a two-part output that renders verified or corroborated current state alongside a clearly-labeled section of human-provided strategy and observations not verifiable from telemetry; such an output can be evaluated for retention of human knowledge clauses and for unsupported state claims, though not, absent a recipient study, for downstream performance benefit.}

\subsection{\texorpdfstring{\bh{Implications of model progress}}{Implications of model progress}}

\bh{Our end-to-end comparison used a specific commercial model at a specific point in time, and base-model capability has advanced since these data were generated. It is plausible that a current frontier model, given the same raw inputs, would close some or all of the estimated-utility gap and reduce hallucination rates; whether it would match the structured pipeline's misinformation profile is an empirical question we have not yet tested, and we are careful not to claim otherwise. We note, however, that several benefits of explicit reconciliation are independent of generator strength: the intermediate representation is auditable, conflicts are inspectable artifacts rather than silent resolutions, provenance survives into the output stage, and behavior is substantially reproducible because everything after each structured LLM output is deterministic. For high-stakes handover, these properties matter even if a stronger model narrows the headline metric differences. Any deployment atop hosted, silently-updated models must also re-validate per model snapshot rather than relying on results obtained under an earlier one.}

\subsection{\texorpdfstring{\bh{What the evaluation metric can and cannot establish}}{What the evaluation metric can and cannot establish}}

\bh{Our information access cost framework is useful for controlled, artifact-level comparison of reports describing the same underlying task state: it is grounded in the task's actual topology and rules, it distinguishes omission from misinformation, and it weights facts by modeled rediscovery cost rather than treating all facts as equally valuable. It is equally important to state what it cannot establish. It does not measure recipient behavior, comprehension, trust, or time-to-resumption; it is task-specific, and its constants (baseline search costs, the need-discovery constant, the misinformation weight) are modeling choices that would require domain-specific re-derivation before use elsewhere; and it credits only state content, so it systematically undervalues strategic guidance. We see its appropriate roles as experimental triage, system debugging, and controlled comparison, not as a substitute for recipient-side validation, particularly in any clinical application.}

\subsection{Limitations and future work}
This study provides a \bh{controlled} investigation into the use of explicit \bh{state} reconciliation techniques alongside LLMs to support task handover; as such, several limitations should be acknowledged\bh{. We enumerate them explicitly, each tied to the claim boundary it imposes:}

\bh{\begin{enumerate}
    \item \textbf{Controlled and simplified domain.} The task was simulated, short, and novel to all participants, lacking professional expertise, real stakes, shared schemas, interactive clarification, and longitudinal shifts. Task novelty likely inflated the value participants placed on experiential guidance; a more experienced worker may report more state, or better calibrate to what the recipient already knows.
    \item \textbf{Designed observability boundary.} Dialogue content was intentionally excluded from telemetry to create a controlled test of source complementarity. This design choice may overstate the benefit of fusing human input relative to richly instrumented domains where less information is human-only.
    \item \textbf{Telemetry treated as authoritative.} Our environment guarantees that logged events are correct; real-world records can be stale, missing, or wrong, which would complicate the conflict-resolution policies that our results support.
    \item \textbf{Task-specific metric assumptions.} Baseline search costs partly reflect search-space cardinality and fixed interaction constants rather than literal path costs, and the need-discovery constant and misinformation weight $\alpha$ are modeling choices. These parameters are applied identically to every condition, and all inferential conclusions rest on the separately reported components rather than the $\alpha$-weighted index; applying the metric in another domain would require re-deriving its parameters for that domain.
    \item \textbf{LLM-dependent extraction and alignment.} Report normalization, existential resolution, and evaluation-time fact extraction all invoke LLMs, so model errors can enter both before generation and during scoring. As described in Section 5, we contain this risk by confining the LLM's evaluative role to surface-form extraction validated against human co-annotation, scoring deterministically against ground truth, and drawing the central contrasts between machine-generated conditions that share the same output template \cite{chen_humans_2024, calderon_alternative_2025}.
    \item \textbf{Strategic knowledge is identified but not valued.} The content analysis shows that human reports carry substantial strategic knowledge, but we do not evaluate its downstream utility; our metrics credit state content only.
\end{enumerate}}

\bh{These limitations scope, rather than undermine, the contribution: within a controlled environment with known ground truth, provenance-aware reconciliation measurably changed what generated handover reports contain and what errors they carry. Future work includes surfacing provenance and conflict records in the rendered report and evaluating their effect on sender and recipient trust and verification behavior; producing a two-part output that preserves human-only strategy alongside verified state; and extending the evaluation to recipients resuming the task from these reports.}

\section{\texorpdfstring{\bh{Conclusion}}{Conclusion}}

\bh{Handover support systems must combine information sources that differ in observability and reliability. In a controlled multitask handover environment with known ground truth, reconciling outgoing workers' reports with system telemetry produced handover reports with greater estimated task-state utility than either source alone. Compared with direct end-to-end LLM synthesis over the same inputs, explicit structured reconciliation maintained comparable estimated utility while incurring substantially less misinformation, and task-aware realization improved the utility--length tradeoff relative to exhaustive reporting. Human-authored reports additionally contained substantial strategic knowledge that state-focused pipelines and metrics do not capture, pointing toward a division of labor in which AI reconstructs and verifies current state while people contribute strategy, intent, and warnings. Together, these findings support provenance-aware state reconciliation as an auditable design pattern for safer AI-assisted handover. They establish artifact-level, estimated task-state benefits; validating those benefits with task-resuming recipients is the natural next step for this line of work.}

\bibliographystyle{ACM-Reference-Format}
\bibliography{references}










\end{document}